\documentstyle[multicol,pre,aps,epsf,epic,eepic,eepicemu]{revtex}
\makeatletter
\def\endtable{%
\global\tableonfalse\global\outertabfalse
{\let\protect\relax\small\vskip2pt\@tablenotes\par}\xdef\@tablenotes{}%
\egroup\vskip 1pc
}%
\def\references{%
\list{\@biblabel{\arabic{enumiv}}}%
{\labelwidth\WidestRefLabelThusFar  \labelsep4pt %
\leftmargin\labelwidth %
\advance\leftmargin\labelsep %
\ifdim\baselinestretch pt>1 pt %
\parsep  4pt\relax %
\else %
\parsep  0pt\relax %
\fi
\itemsep\parsep %
\usecounter{enumiv}%
\let\p@enumiv\@empty
\def\theenumiv{\arabic{enumiv}}%
}%
\let\newblock\relax %
\sloppy\clubpenalty4000\widowpenalty4000
\sfcode`\.=1000\relax
\ifpreprintsty\else\small\fi
}
\makeatother
\begin{document}
\preprint{UGVA-DPT 1998/06}
\draft

\title{Kinetics of ballistic annihilation and branching}
\author{Pierre-Antoine Rey}
\address{Theoretical Physics, Oxford University, 1 Keble Road, Oxford
OX1 3NP, United Kingdom.}
\author{ Michel Droz}
\address{D\'epartement de Physique Th\'eorique, Universit\'e de Gen\`eve,
CH-1211 Gen\`eve 4, Switzerland.}
\author{Jaros\l aw Piasecki}
\address{Institute of Theoretical Physics, Warsaw University, Ho\.za 69, 
Pl-00 681 Warsaw, Poland}

\date{\today}
\maketitle

\begin{abstract}
We consider a one-dimensional model consisting of an assembly of
two-velocity  particles moving freely between collisions. When two
particles meet, they instantaneously annihilate each other and
disappear from the system. Moreover each  moving particle can
spontaneously generate an offspring having the same velocity as its
mother with probability $1-q$. This model is solved analytically in
mean-field approximation and studied by numerical simulations. It is
found that for $q=1/2$ the system exhibits a dynamical phase
transition. For $q<1/2$, the slow dynamics of the system is governed
by the coarsening of clusters of particles having the same velocities,
while for $q>1/2$ the system relaxes rapidly towards its stationary
state characterized by a distribution of small cluster sizes.
\end{abstract}

\begin{multicols}{2}
\section{Introduction}
%%%%%%%%%%%%
Ballistically-controlled reactions provide simple examples of
non-equilibrium systems with complex kinetics and have recently
attracted a lot of
interest~\cite{EF,KS,BRL,R,jarek_uno,jarek_due,jarek_tre,jarek_four,jarek_five}.
They consist of an assembly of particles moving freely between
collisions with given velocities. When two particles meet, they
instantaneously annihilate each other and disappear from the system.

Depending on the initial velocity distribution, two classes of
asymptotic states have been observed in one dimensional systems. In
general, for continuous initial velocity
distribution~\cite{BRL,univer}, as well as for some special case of
discrete velocity distribution (symmetric two-velocity
distribution~\cite{EF,KS,jarek_uno}, or symmetric trimodal velocity
distribution with a sufficiently small fraction of immobile
particles~\cite{jarek_due,jarek_tre}), the steady-state turns out to
be empty and it is approached algebraically in time. The dynamical
exponent characterizing the time decay depends on the initial velocity
distribution and it is still not completely clear how to characterize
the universality classes for this problem~\cite{univer}. On the
contrary, for some discrete velocity distribution, the stationary
state may not be empty, but may contain particles moving all with the
same velocity (for example non-symmetric bimodal velocity
distribution~\cite{EF,jarek_uno} or a trimodal velocity distribution
with more than 25\% of particles initially at
rest~\cite{jarek_due,jarek_tre}). This non-interacting state is
generally approached with an exponentially fast decay.

A richer behavior can be expected in a system with, in opposition to
the ballistic annihilation case, an interacting steady-state. This can
be achieved by constantly bringing new particles in the system by some
suitable mechanism. A possibility is to allow branching processes:
ballistically moving particles can spontaneously generate, with a
given branching rate, some offsprings. Accordingly, one speaks of
ballistic branching-annihilation.

The problem of branching-annihilation has been recently studied in the
framework of a diffusive dynamics~\cite{cardy_uno,cardy_due}. The
simplest example of such a system would be one with a single species
of particle $A$, undergoing diffusive behavior, single--particle
annihilation $A \to \emptyset$, and branching $A \to 2 A$. There is
always a trivial absorbing state, with no particles. For sufficiently
low branching rate, this is the only stationary state, but for larger
values of this rate, another non-trivial `active' stationary state
appears. This stationary state phase transition belongs to the
directed percolation universality class~\cite{dp}. A slightly more
complicated class of model are reaction-diffusion systems with the
underlying reaction processes $2 A \to \emptyset$ and $A \to (m+1) A$,
with $m$ even. It turns out that for these models the critical
exponents are not the ones of directed percolation but belong to a new
universality class~\cite{cardy_uno,cardy_due} characterized by
branching and annihilating walks with an even number of
offsprings. The constraint of local `parity' conservation is the
reason for the existence of this new universality class.

Our aim  here is to study the problem of ballistic
branching-annihilation (BBA) in one dimension for which interesting
new properties can be foreseen. The paper is organized as follows. In
section \ref{sec:model}, the BBA model is defined. The exact dynamical
equations of motion are derived for the one dimensional case. In
section \ref{sec:mf}, the dynamics of the model is studied within a
mean-field like approximation. In particular, the phase diagram of the
steady-state is established in terms of the different parameters of
our model. In this approximation, the steady-state is always
approached exponentially fast. Section \ref{sec:num} is devoted to
numerical simulations of the one dimensional model. It is shown that
fluctuations plays a crucial role. Indeed, as in the mean-field
approximation, a phase transition occurs when the probability that the
offspring takes the velocity of its mother is $q=1/2$; however, for
$q<1/2$ the dynamics is be governed by the coarsening of clusters of
particles having the same velocity, and the system approaches a
completely filled stationary state with a power law decay. For
$q>1/2$, there is no coarsening and the system relaxes rapidly towards
a non-filled stationary state. Finally, the results are discussed in
section \ref{sec:conc}.

\section{The model} \label{sec:model}

We shall first define precisely the BBA model studied and secondly
derive the corresponding  equations of motion.

\subsection{Definition of the model}

We consider a one-dimensional system composed of particles of size
$\sigma$  initially uniformly randomly distributed in space. Moreover,
at $t=0$, the velocities of the particles are random independent
variables distributed with the symmetric bimodal distribution:
\begin{equation}
P(v)=\frac{1}{2}\big[\delta(v-c) + \delta(v+c)   \big]
\end{equation}
The dynamics consists of two mechanisms:
\begin{itemize}
\item{The ballistic annihilation:} Two particles making contact (with
opposite velocities) disappear instantaneously.
\item{The branching:} during the time interval $[t, t+dt]$, the
following branching processes take place:
\begin{enumerate}
\item A particle with coordinates (position and velocity) ($x,~c$)
produces with probability $p(1-q)dt$ a pair of particles with
coordinates  ($x-\sigma,~c$ ) and ($x,~c$).
\item A particle with coordinates  ($x,~c$)  produces with probability
$pqdt$ a pair of particles with coordinates ($x-\sigma,~-c$ ) and
($x,~c$).
\item A particle with coordinates  ($x,~-c$)  produces with
probability $p(1-q)dt$ a pair of particles with coordinates ($x,~-c$ )
and ($x+\sigma,~-c$).
\item A particle with coordinates ($x,~-c$)  produces with probability
$pqdt$ a pair of particles with coordinates ($x,~-c$ ) and
($x+\sigma,~c$).
\end{enumerate}  
\end{itemize}
(the particular choice of the position of the newly created particle
has been made in order that, independently of its velocity, a child
cannot collide with its mother at birth.) Thus the parameter $0
\le p \le \infty$ characterizes the overall branching rate, while the
parameter $0 \le q \le 1$ characterizes the probability that the
offspring has a velocity opposed to the one of its mother. The
particular case $p=0$ corresponds to the pure ballistic annihilation
problem previously studied~\cite{EF,KS,jarek_uno,jarek_due}.

\subsection{Exact equations of motion}

We can  now derive the equations of motion describing the dynamics of
the system. In the particular case $p=0$, a kinetic equation for the
two-particle conditional distribution of nearest neighbors was
derived as a rigorous consequence of the dynamics of ballistic
annihilation~\cite{jarek_uno,jarek_due}. This equation completely
described the evolution of the system when initially higher order
conditional distributions factorized into products of two-particle
ones. It was then possible to extract exactly and analytically the
long time behavior of the particle density for several velocity
distributions. Unfortunately, this property is no longer valid in the
case with branching. Having not been able to find an observable in
which one is able to reproduce this exact closure, one has to face the
usual problem of dealing with a complete hierarchy of coupled
equations~\cite{bbgky}. It seems thus hopeless to find an exact
analytical solution to these equations. Accordingly we shall only
write the equation for the one-particle density distribution
$\rho_1(x,v;t)$. In section \ref{sec:mf}, this equation will be solved
using a mean-field approximation.
\end{multicols}
\vspace{-4.8mm}
\noindent\rule{20.5pc}{0.25pt}\rule{0.25pt}{5pt}

A careful bookkeeping of the possible dynamical processes leads to
the following equations:
\begin{eqnarray}
(\partial_t + c \partial_x)\rho_1(x,c;t)
&=& -2c\rho_2(x,c;x+\sigma,-c;t) \nonumber \\
&+& pq\Big[\rho_1(x-\sigma,-c;t) - \sum_{v=\pm c}\int_0^{\sigma}dy\,
           \rho_2(x-\sigma,-c;x+y,v,t) \Big] \nonumber \\
&+& p(1-q)\Big[\rho_1(x+\sigma,c;t) - \sum_{v=\pm c}\int_0^{\sigma}dy\,
               \rho_2(x-y,v;x+\sigma,c;t) \Big] , \label{r1}
\end{eqnarray}
and
\begin{eqnarray}
(\partial_t - c \partial_x)\rho_1(x,-c;t)
&=& -2c\rho_2(x,c;x+\sigma,-c;t)  \nonumber \\
&+& pq\Big[\rho_1(x+\sigma,-c;t) - \sum_{v=\pm c}\int_0^{\sigma}dy\,
           \rho_2(x-y,v;x+\sigma,c;t) \Big]  \nonumber \\
&+& p(1-q)\Big[\rho_1(x-\sigma,-c;t) - \sum_{v=\pm c}\int_0^{\sigma}dy\,
               \rho_2(x-\sigma,-c;x+y,v;t) \Big]   \label{r2}
\end{eqnarray}
\hspace*{22pc}\rule[-4.5pt]{0.25pt}{5pt}\rule{20.5pc}{0.25pt}%
\vspace*{-8pt}
\begin{multicols}{2}
where $\rho_2(x_1,v_1;x_2,v_2;t)$ is the joint two-particle density to
find a particle in the state $(x_1,v_1)$ simultaneously with another
in the state $(x_2,v_2)$ at time $t$. 

The right-hand side of equation (\ref{r1}) can be interpreted in the
following way: the first term describes the annihilation of a particle
$(x,c)$ with a particle of opposite velocity. It is given by the
product of the density of a collision configuration
[$\sigma\rho_2(x,c;x+\sigma,-c,t)$] with the frequency of such an
encounter ($2c/\sigma$). The second term describes the branching of a
particle of velocity $-c$, at position $x-\sigma$, giving birth to a
particle of velocity $+c$ at position $x$. This is only possible if no
other particles are present in the interval $[x,x+\sigma]$ (otherwise
there will be an overlap between two particles) and it happens with
the rate $pq$. Finally, the third term describes the creation with
rate $p(1-q)$ of a particle whose mother have the same velocity. The
same restriction as in the previous case applies.

One can in principle write the equation of motion for $\rho_2$ along
the same lines. However, we shall not give here this cumbersome
equation, as we are not going to use it.

For simplicity, we shall only consider spatially homogeneous system.
We can thus write $\rho_1(x,v;t)=\rho_1(v,t)$ and
$\rho_2(x_1,v_1;x_2,v_2;t) =\rho_2(x_1-x_2,v_1;0,v_2;t)$. Introducing
then the observable $\Psi(t) \equiv \rho_1(c,t) - \rho_1(-c,t)$, one
easily shows that it is an exactly conserved quantity when
$q=1/2$. This feature reflects the particular choice of rule, which
are precisely symmetric when $q=1/2$. As a consequence, one expects
our model to exhibit a particular behavior at this point.

\section{Mean-field analysis} \label{sec:mf}

A first attempt to obtain information about our model is to apply a
mean-field approximation on equations~(\ref{r1}) and~(\ref{r2}). One then
assumes the following factorization:
\begin{eqnarray}
\rho_2(x_1,v_1;x_2,v_2;t)
&=& \rho_1(x_1,v_1;t)\rho_1(x_2,v_2;t) \nonumber \\
&=& \rho_1(v_1,t)\rho_1(v_2,t),
\label{mf}
\end{eqnarray}
(the last equality holds for a spatially homogeneous system).

It is then suitable to introduce in addition to the variable $\Psi$
the second  variable:
\begin{equation} 
\Phi(t) \equiv \rho_1(c,t) + \rho_1(-c,t), \label{sum}
\end{equation}
Equations~(\ref{r1}) and~(\ref{r2}) lead to
\begin{equation} 
\frac{d\Phi}{dt}
 =  p(1-\sigma \Phi)\Phi - c(\Phi^2-\Psi^2),\label{eqm1}
\end{equation} 
and
\begin{equation} 
\frac{d\Psi}{dt}
 =  p(1-2q)(1-\sigma \Phi) \Psi. \label{eqm2}
\end{equation} 
The formal solution of this last equation is
\begin{equation} 
\Psi(t)
 =  \Psi(0)\exp
    \Bigl[p(1-2q)\Bigl(t-\sigma \int_0^t d\tau\,\Phi(\tau)\Bigr)\Bigr],
\label{phi}
\end{equation} 
As before, one sees that the value $q=1/2$ plays a special
role. Indeed, two regimes have to be distinguished:
\begin{enumerate}
\item{For $0 \le q <1/2$:} the exponential term diverges unless
$(1-\sigma \Phi) \to 0$ as $t \to \infty$. Thus a possible stationary
solution is
\begin{equation} 
\Phi_s = \frac{1}{\sigma}, \quad
{\Psi_s}^2 = \frac{1}{\sigma^2}, \label{statio1}
\end{equation}
In the particular case $q=0$, the time dependent solution can be obtained explicitly as shown in Appendix. For $t \to \infty$, one recovers the above stationary solution.

\item{For $1/2 < q \le 1$:} in this case, a possible stationary
solution is
\begin{equation} 
\Psi_s = 0, \quad
\Phi_s= \frac{1}{\sigma}{(1+c/p\sigma)}^{-1},\label{statio2}
\end{equation}
\end{enumerate}
Is is straightforward to verify that the above stationary solutions are
stable and are approached exponentially in time.

Moreover, when $q=1/2$ the complete time dependent solution can be
obtained. From equation~(\ref{eqm2}), one indeed finds $\Psi(t)={\rm
const}=\Psi_0$, and thus equation~(\ref{eqm1}), becomes
\begin{equation} 
\frac{d\Phi}{dt}
 =  p \Phi - (c+p \sigma)\Phi^2+c \Psi_0^2, \label{eqpart}
\end{equation} 
whose solution reads
\begin{equation} 
\Phi(t)
 =  \frac{p}{2(c+p\sigma)}
 +  \frac{\gamma A \cosh(At)+A(c+p\sigma)\sinh(At)}
         {A\cosh(At)+\gamma(c+p\sigma)\sinh(At)},\label{timedep}
\end{equation}
with $A=p^2/4+c(c+p\sigma)\Psi_0^2$ and
$\gamma=\Phi(0)-p/[2(c+p\sigma)]$. The stationary state is then given
by $\Psi_s=\Psi_0$ and
\begin{equation} 
\Phi_s(q=1/2)
 =  \Bigl(p+\sqrt{p^2+4c(c+p\sigma)\Psi_0^2}\Bigr)\Big/(c+p\sigma).
\label{stat22}
\end{equation}
Here again, one sees from equation~(\ref{timedep}) that the steady
state is approached in an exponential way. As already noted, $\Psi(t)$
is an exactly conserved quantity for $q=1/2$.

\begin{figure}
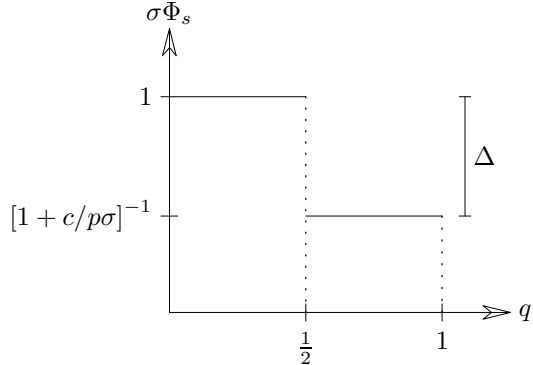

{\samepage\columnwidth20.5pc
\centerline{\input mf.eepicemu}
\caption{Mean-field phase diagram: the stationary value of the
         averaged density $\Phi_s$ is plotted against $q$ for a fixed
         value of $p$.\label{mfphase}}
}
\end{figure}

The mean-field stationary phase diagram is shown in Fig.\
\ref{mfphase}. The stationary value $\Phi_s$ is plotted against $q$
for a fixed value of $p \not= 0$. The interesting feature is the
presence of a gap $\Delta(p)$ for $q=1/2$ given by
\begin{equation} 
\Delta(p) = \frac{1}{\sigma}\Big(1-\frac{1}{1+c/p\sigma}\Big).
\end{equation} 
$\Delta(p)$ decreases as $p$ increases. When $q < 1/2$,
$\Phi_s=1/\sigma$ for all values of $p$ (completely filled state),
while for $q > 1/2$, $\Phi_s$ increases monotonically with $p$. The
dependence is linear for small $p$, but $\Phi_s \to 1/\sigma$ when $p
\to \infty$.

\section{Numerical simulations} \label{sec:num}

In view of the situation when $p=0$~\cite{EF,KS,jarek_uno}, one can
anticipate that the fluctuations will also play an important role in
the case with branching. One way to deal with the complete problem,
including fluctuations, is to perform numerical simulations.

The simulations were performed for a one-dimensional periodic lattice
with typically $2^{17}$ sites. The velocity of each particle was
drawn from a symmetric bimodal distribution. However, on
computational grounds, the particle velocities were chosen to be
$(0,c')$ (with $c'>0$). The results for our model defined in section
\ref{sec:model} can be recovered by performing a simple Galilean
transformation and putting $c=c'/2$. The particle size $\sigma$ is the
lattice spacing, and the discretized time step is given by
$\tau=\sigma/c'$.

The algorithm used to simulate the dynamics is the following. During
one time step $\tau$, the three following processes occur sequentially:
\begin{enumerate}
\item{Ballistic motion:} independently of the occupation state of
the sites, the particles with velocity $c'$ move one site to the
right.
\item{Annihilation:} two particles located on the same site disappear.
\item{Branching:} for each remaining particle, one draws two random
numbers, $r_p$ and $r_q$, uniformly distributed in the interval
$[0,1]$. One offspring particle is added to the left (right) nearest
neighbor of a particle with velocity $c'$ ($0$) if the site is empty
and if $r_p$ is the less than a given value $\tilde p$. Hence, $\tilde
p$ is the probability of branching. This new particle takes the
velocity of its mother with a certain probability $1-q$, i.e. if
$r_q>q$ (and the other velocity otherwise). If two particles are
created on the same site (thus born from two different mothers), they
annihilate instantaneously.
\end{enumerate}
For each of the above different steps, the sites were updated
simultaneously. The simulations were run on a Connection Machine
CM-200 and the data averaged over $10$ independent realizations. The
mean initial density for all the simulations was $0.5$, with, in
average, the same densities of both kinds of particles. We have also
shown that our results obtained for lattice of $2^{17}$ sites were
free of finite size effects.

Note that when a particle branches, it can create at most one particle
during a time step $\tau$. As a consequence, this limit the value of $p$
that can be explored through the simulations. Indeed, the branching
rate $p$ is related to $\tilde p$ via
\begin{equation}
\tilde p = p \tau.
\end{equation}
Thus using the definition of $\tau$ and $c'$, one finds
\begin{equation}
\frac{p\sigma}{c} = 2\tilde p.
\end{equation}
$\tilde p$ being a probability, the adimensional branching rate
$p\sigma/c$ can only take values between $0$ and $2$.

We can now discuss the numerical data obtained using the above
algorithm. Two kinds of quantities have been investigated: first, the
time dependent density with particular emphasis on the stationary
states and the way these stationary states are approached; second, a
more microscopic quantity, namely the time dependent cluster size
distribution $P(\ell,t)$ in the system and some of its moments. These
quantities are well suited to describe the coarsening process present
in the system.

\begin{figure}
\epsfxsize=7.5cm
{\samepage\columnwidth20.5pc
\centerline{\epsfbox[40 60 550 720]{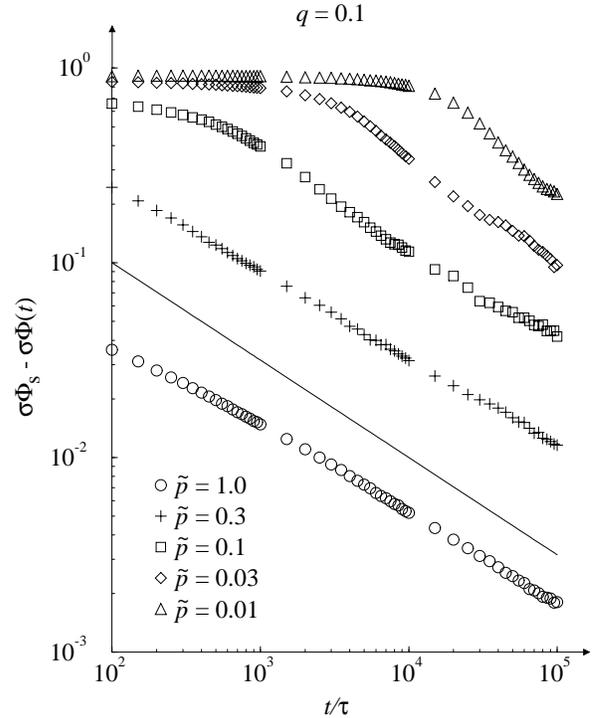}}
\caption{$\sigma\Phi_s-\sigma\Phi(t)$ versus $t$ in a double
         logarithmic scale, for $q=0.1$ and several values of $\tilde
         p$. For comparison, the full line represents $t^{-1/2}$. This
         decay establishes after a crossover time which behaves as
         $\tau/\tilde p$.\label{appless}}
}
\end{figure}

\begin{figure}
\epsfxsize=7.5cm
{\samepage\columnwidth20.5pc
\centerline{\epsfbox[40 60 550 720]{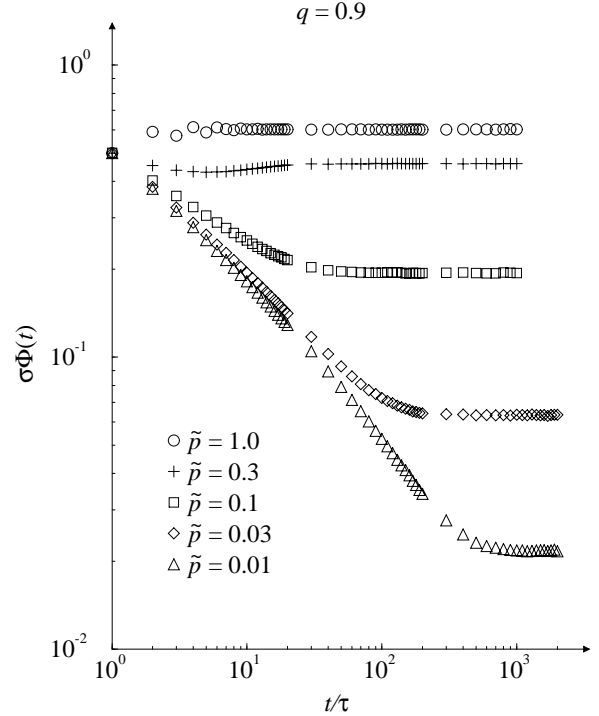}}
\caption{Time evolution of the particle density $\sigma\Phi(t)$ as a
         function of time $t$ for $q=0.9$ and several values of
         $\tilde p$. The stationary state is reached after a time of
         order $10\tau/\tilde p$.\label{phimore}}
}
\end{figure}

\begin{figure}
\epsfxsize=7.5cm
{\samepage\columnwidth20.5pc
\centerline{\epsfbox[40 60 550 720]{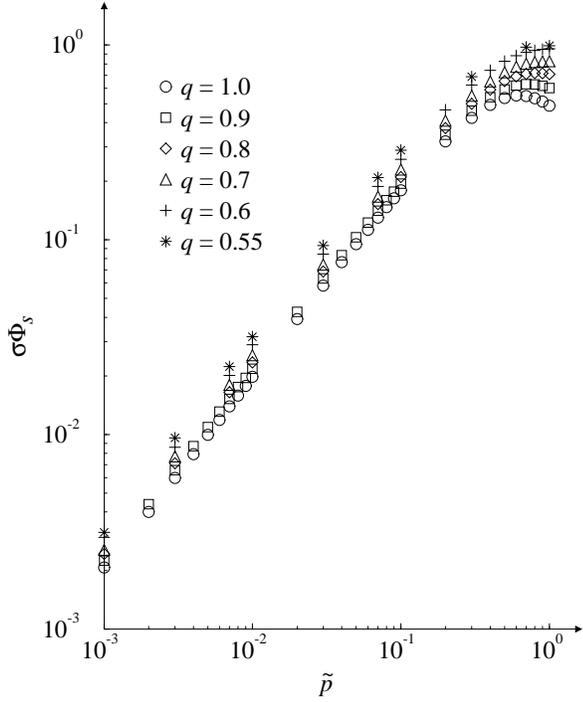}}
\caption{The stationary values of the averaged density $\sigma\Phi_s$
         is plotted against $\tilde p$, for several values of $q>1/2$,
         obtained by numerical simulations.\label{statmore}}
}
\end{figure}

\begin{figure}
\epsfxsize=7.5cm
{\samepage\columnwidth20.5pc
\centerline{\epsfbox[40 60 550 720]{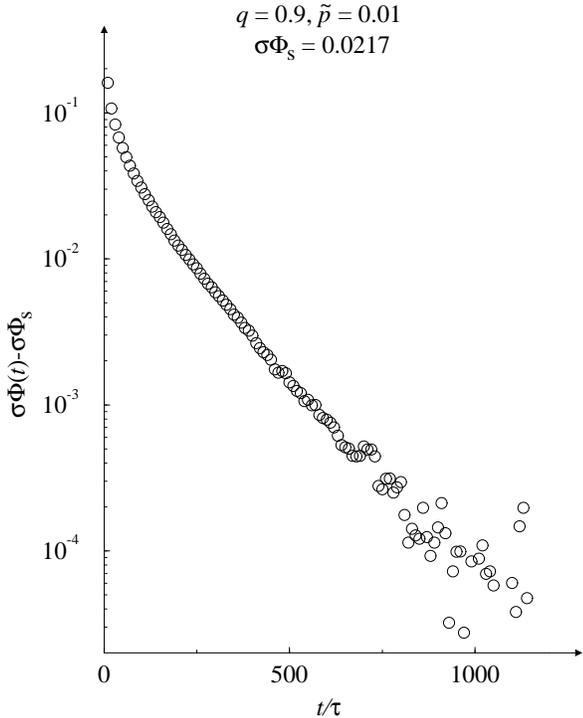}}
\caption{Semi-logarithmic plot of $\sigma\Phi(t)-\sigma\Phi_s$ versus
         $t$ for $q=0.9$ and $\tilde p=0.01$. The exponential approach
         towards the steady state establishes for $t/\tau \simeq
         250$.\label{appmore}}
}
\end{figure}

\begin{figure}
\epsfxsize=7.5cm
{\samepage\columnwidth20.5pc
\centerline{\epsfbox[40 60 550 720]{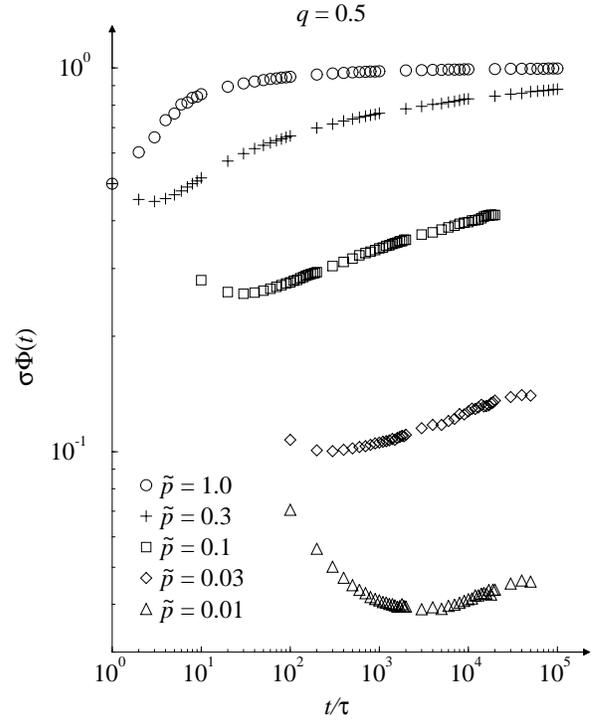}}
\caption{Time evolution of the particle density $\sigma\Phi(t)$ as a
         function of time $t$ for $q=0.5$ and several values of
         $\tilde p$. For small values of $\tilde p$ (less than 0.3),
         we are unable to extract the steady-state density, for CPU
         reasons.\label{phiequal}}
}
\end{figure}

As in the mean-field approach and as expected from the last remark of
section \ref{sec:model}, the value $q=1/2$ turns out to play a
particular role and three regimes have to be distinguished.
\begin{enumerate}
\item{For $0 \le q <1/2$:} The time evolution of the particle density
$\Phi(t)$ is shown in Fig.\ \ref{philess} for several values of
$\tilde p$. Clearly, the system reaches a stationary state
$\Phi_s=1/\sigma$ in agreement with the mean-field
prediction. However, as shown in Fig.\ \ref{appless}, the stationary
state is approached as $\Phi_s-\Phi(t) \sim t^{-1/2}$. This power law
establishes after a crossover time roughly proportional to $1/p$.
\begin{figure}
\epsfxsize=7.5cm
{\samepage\columnwidth20.5pc
\centerline{\epsfbox[40 60 550 720]{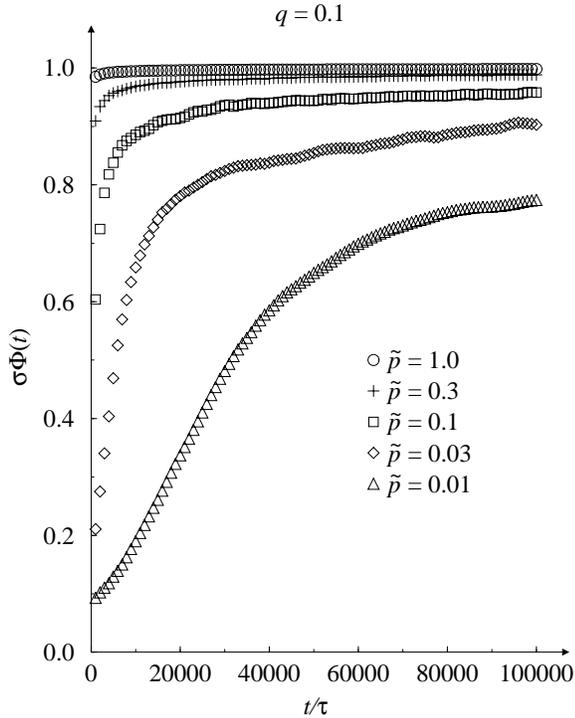}}
\caption{Time evolution of the particle density $\sigma\Phi(t)$ as a
         function of time $t$ for $q=0.1$ and several values of
         $\tilde p$.\label{philess}}
}
\end{figure}
\item{For $1/2 < q \le 1$:} The time evolution of the particle density
$\Phi(t)$ is shown in Fig.\ \ref{phimore} for several values of
$\tilde p$. As depicted on Fig.\ \ref{statmore}, the stationary value
of the density depends both on $\tilde p$ and $q$. For $\tilde p <
0.1$, it is well fitted by
\begin{equation}
\Phi_s(\tilde p,q) \approx {\tilde p} \exp(0.55/q), \label{smallp}
\end{equation}
Moreover, for $\tilde p$ large enough $\Phi_s$ is not increasing
monotonically as a function of $\tilde p$, but $\Phi_s$ exhibits a
maximum and then slightly decreases as $\tilde p$ increases. As shown
in Fig.\ \ref{appmore}, the stationary state is approached in an
exponential way according to $\Phi_s-\Phi(t) \sim \exp(-A\tilde pt)$,
where $A$ may depend on $q$.
\item{} The limit case $q=1/2$ is more difficult to investigate due to
the slow decay towards the stationary state. In fact for $\tilde p >
0.3$, there are evidences that the stationary state is completely
filled, i.e. $\Phi_s=1/\sigma$. For smaller values of $\tilde p$ the
simulations do not allow us to draw any conclusions, as shown in Fig.\
\ref{phiequal}.

Nevertheless, for $q <1/2$, one has $\Phi_s=1/\sigma$, for all
values of $\tilde p$ while for $q>1/2$, equation~(\ref{smallp}) shows
that, at least for small $\tilde p$, $\Phi_s \not= 1/\sigma$. Thus for
small $\tilde p$, $\Phi_s$ has a jump at $q=1/2$, and we believe that
such a jump will be present for all finite values of $p$.
\end{enumerate} 
%\vspace*{-5mm}

\begin{figure}
\epsfxsize=7.5cm
\centerline{\epsfbox{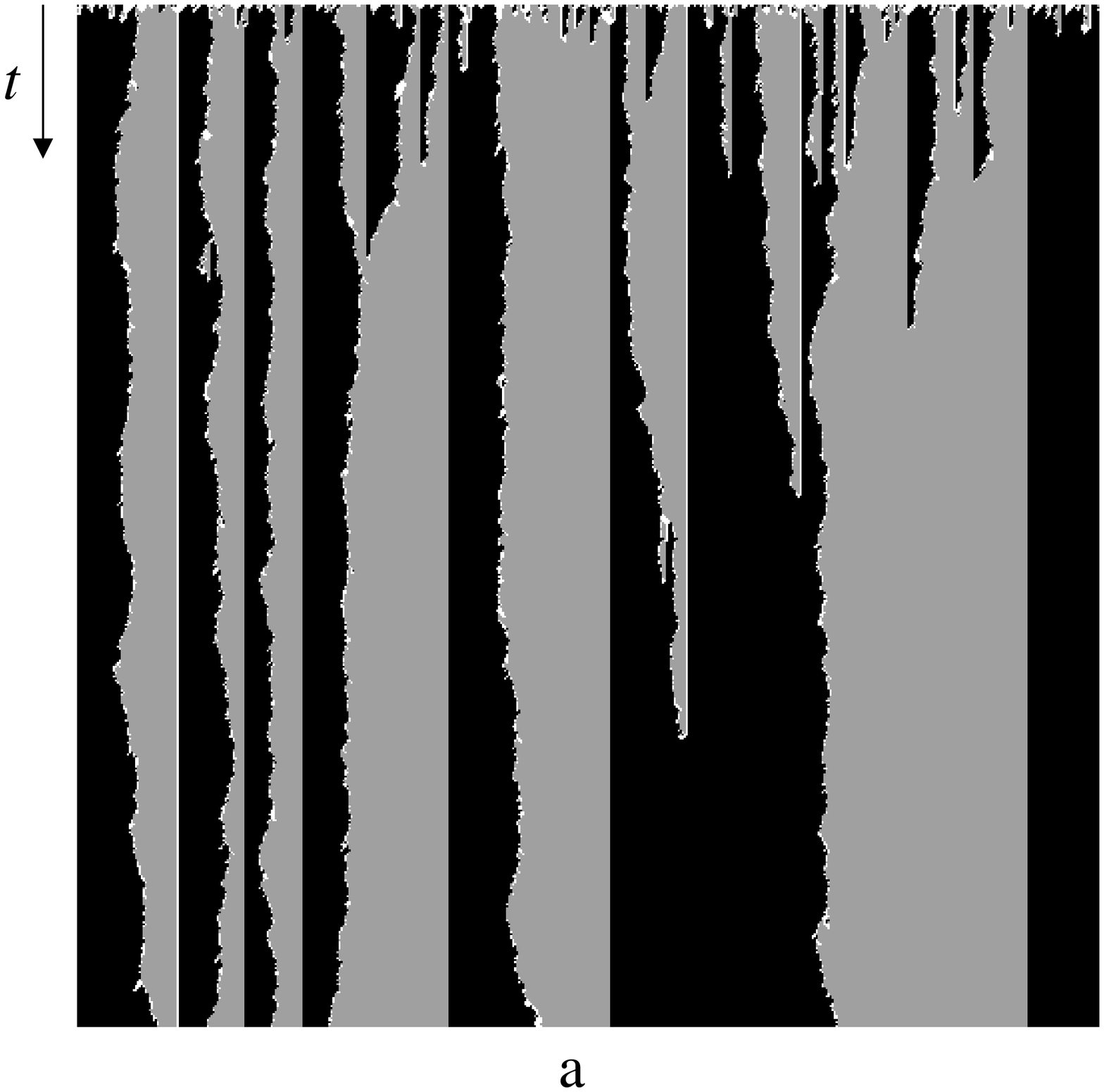}\hspace{1cm}%
\epsfxsize=7.5cm\epsfbox{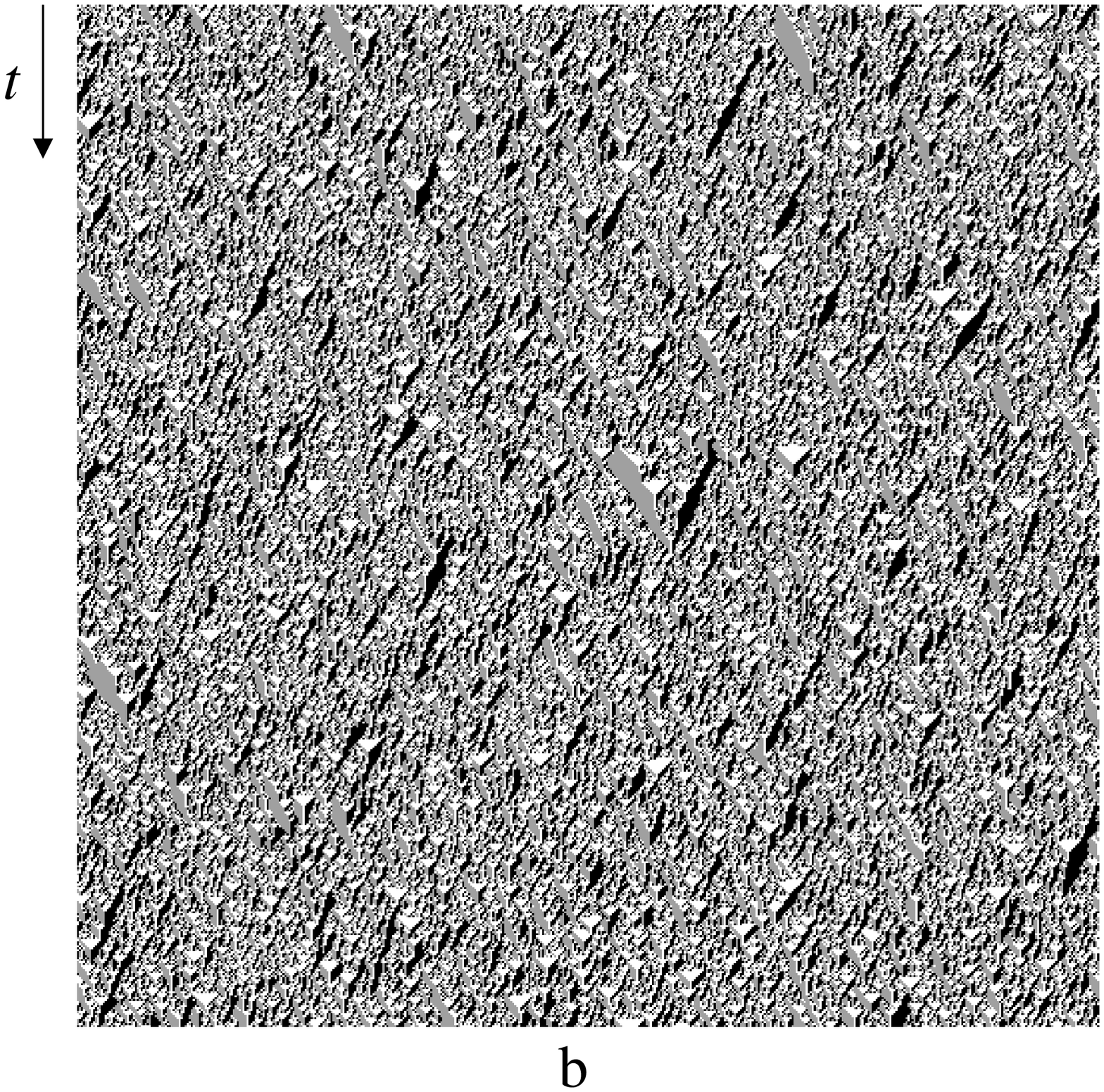}}

\vspace{3mm}
\caption{Time evolution (vertical axes) of the configurations for a
chain of 512 sites (the initial density is approximately one half) and
for 1024 time iterations. The white pixels indicate sites without
particle, the grey ones, sites with a particle towards the right and
the black ones, sites with a particle moving towards the left. Fig.\ a
is for $\tilde p=0.7$, $q=0.9$ while Fig.\ b is for $\tilde p=0.7$,
$q=0.1$. \label{snapshots}}
\end{figure}

We can now consider the properties of the clusters present in the
system at a given time. The qualitative situation is well illustrated
by the two  snapshots in Fig.\ \ref{snapshots}. They represent the
time evolution of a $512$-site system during 1024
iterations. Moreover, a change of reference frame has been performed
such that the particle velocities appears to be $\pm c$. Depending on
$q$, one observes totally different pictures. In the case $\tilde
p=0.7, q=0.1$ (Fig.\ \ref{snapshots}a), large clusters (of similar
particles) are present. They are separated by two types of interfaces:
vertical ones (which are stable) and rough ones. The dynamics of the
system is totally governed by the random walks of the rough
interfaces. During the time evolution, one rough interface may collide
with a stable interface leading to the coalescence of two clusters
into a large one. In the case $\tilde p=0.7, q=0.9$ (Fig.\
\ref{snapshots}b), the sizes of the clusters are rather small and
there is no stable interfaces. The dynamics is of a different type.

A more quantitative description is given by the investigation of the
time dependent cluster size distribution $P(\ell,t)$. In the domain $0
\le q <1/2$, where coarsening is observed, one expects~\cite{frach} that
$P(\ell,t)$ will obey to a scaling form:
\begin{equation}
P(\ell,t) \sim t^{-\alpha} \Pi(\ell t^{-\beta}). \label{scaling}
\end{equation}
\end{multicols}
\begin{multicols}{2}
In Fig.\ \ref{collapse}, we plot the scaling function obtained by the
collapse of the data for $\tilde p=0.7, q=0.1$, with $\alpha=1$ and
$\beta=0.5$. Although the plot is very noisy, one still notes that the
scaling function $\Pi(z)$ has a very particular shape, with a sharp
maximum at $z=z_{max}$. The value of $z_{max}$ increases slowly with
$q$, going from $0.4$ for $q=0.1$ to $1.2$ for $q=0.4$.

\begin{figure}
\epsfxsize=7.5cm
{\samepage\columnwidth20.5pc
\centerline{\epsfbox[40 60 550 720]{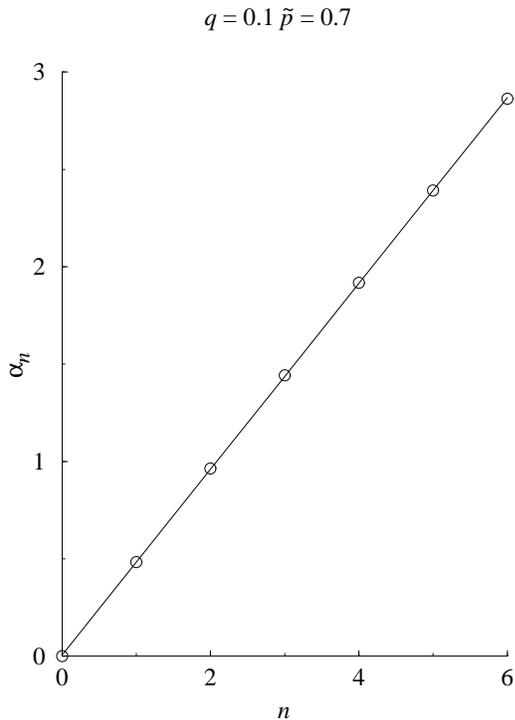}}
\caption{Exponent $\alpha_n$ (open circles) of the $n$-th moment of the
cluster distribution function for $n=0,\ldots,6$ and $\tilde
p=0.7, q=0.1$. The line is the fit $\alpha_n=0.01+0.48 n$.\label{regress}}
}
\end{figure}

A better way to extract the exponents $\alpha$ and $\beta$ is to
consider the $n$-th order moments of the distribution defined as:
\begin{equation}
\langle\ell^n\rangle
 =  \frac{\int_{\sigma}^{\infty} d\ell\,\ell^n P(\ell,t)}
         {\int_{\sigma}^{\infty} d\ell\,P(\ell,t)}
\end{equation}
which according to the scaling form given by equation~(\ref{scaling}),
should behave as:
\begin{equation}
\langle\ell^n\rangle \sim t^{\alpha_n} = t^{n \beta}, \label{scalexp}
\end{equation}
while
\begin{equation}
\int_{\sigma}^{\infty} d\ell\,P(\ell,t)\sim t^{-\alpha + \beta}
\end{equation}
Thus, the two above relations allow us to determine the exponents
$\alpha$ and $\beta$. The values of $\alpha_n$ for $n=1,\dots,6$ are
shown on Fig.\ \ref{regress} for $p=0.7$, $q=0.1$. A good fit is
obtained for $\beta=0.48\pm 0.02$ and $\alpha=0.96\pm 0.04$, in very
good agreement with our collapsed plot. By repeating our analysis for
other values of $q$ (namely, 0.2, 0.3 and 0.4), the same values for the
exponents fit reasonably well the data.

\begin{figure}
\epsfxsize=7.5cm
{\samepage\columnwidth20.5pc
\centerline{\epsfbox[40 60 550 720]{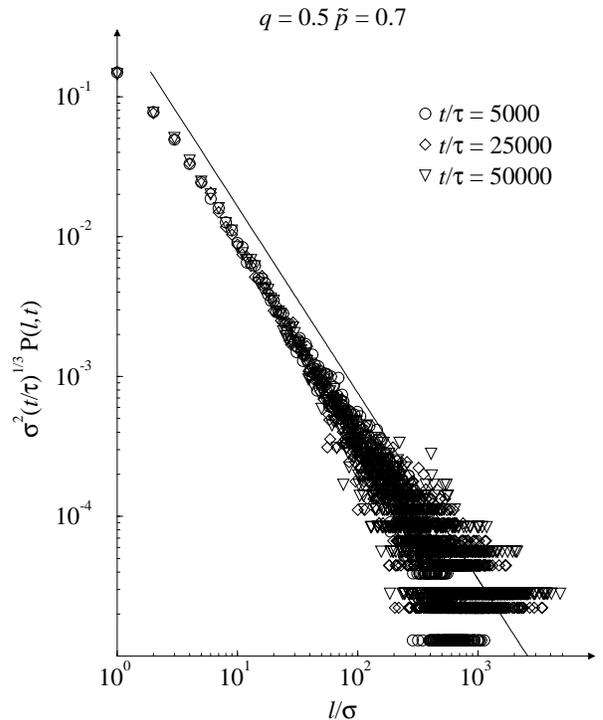}}
\caption{Scaling form of the cluster sizes distribution for $\tilde
p=0.7, q=0.5$. $P(\ell,t) t^{1/3}$ is plotted versus $\ell$ in a
double logarithmic scale. The full line represents
$\ell^{-4/3}$.\label{colequal}}
}
\end{figure}

For $q=1/2$, the different moments of the cluster size distribution
are $\alpha_1=0.33$, $\alpha_2=0.96$, $\alpha_3=1.60$,
$\alpha_4=2.22$, $\alpha_5=2.82$ and $\alpha_6=3.40$. These exponents
are of the form $\alpha_n=-0.26 +0.61n$ which is not compatible with
the relation~(\ref{scalexp}). This probably shows that the simulations
have not yet reached the true asymptotic regime. Moreover, as shown on
Fig.\ \ref{colequal}, $P(\ell,t)$ is of the form:
\begin{figure}
\epsfxsize=7.5cm
{\samepage\columnwidth20.5pc
\centerline{\epsfbox[40 60 550 720]{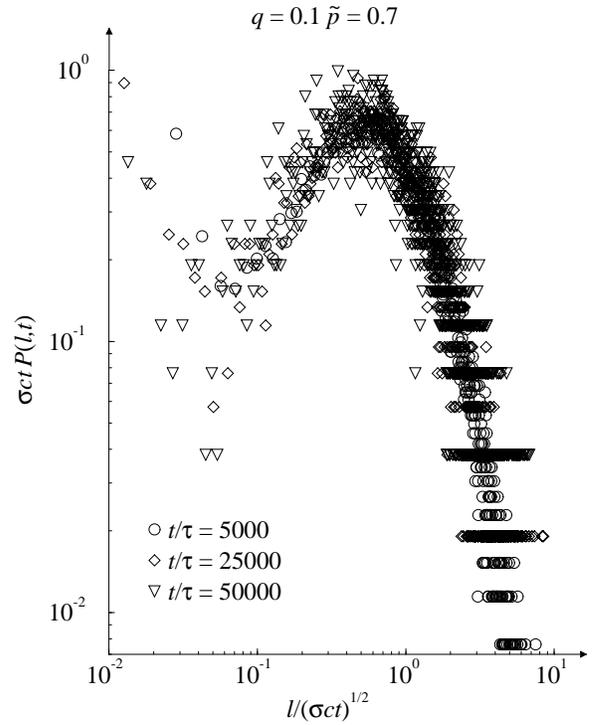}}
\caption{Scaling form of the cluster sizes distribution for $\tilde
p=0.7, q=0.1$. $P(\ell,t)t^{\alpha}$ is plotted versus $\ell
t^{-\beta}$ for $\alpha=1$ and $\beta=0.5$.\label{collapse}}
}
\end{figure}
\begin{equation}
P(\ell,t) \sim t^{-1/3}\ell^{-4/3}, \label{pow}
\end{equation}
over two decades in the variable $\ell$. Note that equation~(\ref{pow})
cannot be valid for arbitrary large $\ell$, because the moments of
$P(\ell,t)$ diverge with the upper limit of integration.

Finally, in the domain $1/2 < q \le 1$, where no coarsening is
observed, the system approaches very  rapidly its stationary state and
no dynamical scaling has been found for the cluster
distribution. However, in the stationary state, the cluster
distribution takes the form:
\begin{equation}
P(\ell) =C_1 \exp(-C_2 \ell)
\end{equation}
where $C_1$ and $C_2$ are two constants.

\section{Interpretation of the results and conclusions} \label{sec:conc}

The first interesting point is the particular role played by the value
$q=1/2$. As already mentioned in section \ref{sec:model}, for $q=1/2$,
one notes the presence of an extra conservation law in the system. The
difference between the average local density of particles with
positive and negative velocities is strictly zero. It is well known
that conservation laws has a great influence on the dynamics of
non-equilibrium statistical systems. Accordingly, one may expect that
the dynamics in $q=1/2$ is particular.

In view of the scaling properties of the problem it may be useful to
think about it in terms of  dynamical renormalization group. Based on
the results of both mean-field approximation and the numerical
simulations, one is lead to conjecture the presence of three fixed
points in this system. An unstable ``critical'' fixed point at $q=1/2$
and two attractive fixed points at $q=0$ and $q=1$.

When $q<1/2$, the branching processes favors the apparition of pair of
consecutive particles with the same velocities and the dynamics is
governed by the attractive fixed point at $q=0$. Large particle
clusters with opposite velocities are formed during the time evolution
and two kinds of interfaces are present into the system (see Fig.\
\ref{snapshots}a). First, let us consider the interface between two
clusters of colliding particles and call this type of interface
$I_1$. Such interface has a very long life. Indeed, the probability
that a vacancy presents at one of the extremity of a cluster of size
$L$ traverses the cluster and perturbates the interface is of the
order of $(1-p)^L$. Thus, an interface $I_1$ is very stable in the
long time limit where the system is made up of large clusters. The
second type of interface, called $I_2$, separates non-colliding
clusters. Thus it has not necessarily a one site extension, but it can
be wider. Hence, its behavior is more subtle. Three different regimes
may be considered. The simplest case to discuss is when
$p\sigma/c>1$. In this case, the interface $I_2$ is typically formed
by only one empty lattice site, whose dynamic is diffusive. Indeed,
one can show that both boundaries of an interface $I_2$ perform a
Brownian motion. Moreover, when $p\sigma/c>1$, this random walk is
biased, so that both boundaries tend to come closer together. For
sufficiently long time, the initial gap separating two non-colliding
clusters will shrink to one single site, which will perform a random
walk. Eventually, this hole will encounter an $I_1$ interface,
permitting the coalescence of two clusters into a larger one. The
random walk aspect of this dynamics is responsible for the slow
approach towards the stationary state (in $t^{-1/2}$) observed in the
simulations. When $p\sigma/c=1$, the boundaries of an interface $I_2$
both perform an unbiased random walk. Accordingly, the initial gap
between two non-colliding clusters will not, on average,
vary. However, because of the BBA dynamics, this gap will eventually
shrink to a single site, either through a creation of cluster inside
the gap when $q\neq0$, or through the coalescence of two
interfaces. Thus the previous argument holds. Finally, when
$p\sigma/c<1$, the situation is similar: although the boundaries of
$I_2$ perform a biased random walk which tends to increase the
separation between the two non-colliding clusters, the coalescence of
two interfaces or a creation of a new cluster inside the gap (if
$q\neq0$) will fill up this space in a more efficient way.

Eventually, the stationary state is completely filled, only one cluster
remains and the annihilation process do not act anymore. For values of
$q$ not too far from $1/2$, this asymptotic behavior will shows up
only for very long times. Accordingly, the results of the (finite time)
numerical simulations may still be  affected by the properties of the
critical fixed point at $q=1/2$, and the dynamics will exhibit some
crossover behavior.

In the situation $q>1/2$, a majority of pairs of particles with
opposite velocities are created during branching.  Due to the
annihilation processes, those particles will prevent the formation of
large clusters of particles. One may anticipate that the long time
dynamics is governed by the other attractive fixed point corresponding
to $q=1$. The dynamics is no longer governed by coarsening mechanism
but only by the dynamics of small clusters, hence the fast
(exponential like) relaxation occurs. Depending upon the value of $p$,
there is a more or less important fraction of empty sites (or holes)
into the system. The presence of these two different dynamical regimes
explains the jump observed in the stationary density at $q=1/2$.

This paper shows once again, that the mean-field results generally do
not hold for low dimensional systems. Whereas the mean-field
approximation predict the exact critical value for $q$ (because the
mean-field equation for $\Psi$ is exact when $q=1/2$) and the right
stationary value of the density when $q<1/2$, it is unable to give
satisfactory results for the density stationary value for $q>1/2$,
(see Figs \ref{mfphase} and \ref{statmore}). Unsurprisingly, the
mean-field approximation is also unable to predict the power law
approach to the stationary state when $q<1/2$, which is obviously
governed by fluctuations. More surprisingly, its prediction of an
exponentially fast approach towards the steady state when $q>1/2$ is
(qualitatively) well verified. However, to better understand this
problem, it would be useful to be able to find an exact analytical
solution at least for the three fixed point cases ($q=0, 1/2$ and $1$
for arbitrary values of $p$) as a support to the above qualitative
picture. Unfortunately, we were not able until now to find such exact
solutions.

In conclusion, one sees that this simple BBA problem with one
offspring exhibit already a very rich behavior. The case with two or
more offsprings is a completely open question.

\section*{Acknowledgments}
Works partially supported by the Swiss National Science Foundation
(M.D). Two of us (P.-A.R. and J.P.) acknowledge the hospitality of Department
of Theoretical Physics of the University of Geneva were part of this
work was done. P.-A.R. is supported by the Swiss National Science
Foundation and J.P. acknowledges the financial support by KBN
(Committee for Scientific Research, Poland) grant 2 P03 B 035 12.

{\appendix
\section{}
In this Appendix, we give an explicit solution to the mean-field
equations~(\ref{sum}) and~(\ref{eqm1}) for the case
$q=0$. Equations~(\ref{sum}) and~(\ref{eqm1}) lead to
\begin{equation}
\Psi\,\frac{d\Phi}{dt} - \Phi\,\frac{d\Psi}{dt}
 =  -c\Psi(\Phi^2-\Psi^2). \label{ap1}
\end{equation}
Multiplying eq.~(\ref{ap1}) by $\Psi^{-2}$, and introducing
$\chi(t)=\Phi(t)/\Psi(t)$, eq.~(\ref{ap1}) becomes
\begin{equation}
\frac{d\chi}{dt} = -c\Psi(\chi^2 -1),
\end{equation}
whose solution is
\begin{equation}
\chi(t)
 =  \frac{1+\chi(0)+[\chi(0)-1]\exp
          \bigl(-c\int_0^t \Psi(\tau) d\tau\bigr)}
         {1+\chi(0)-[\chi(0)-1]\exp
          \bigl(-c\int_0^t \Psi(\tau) d\tau\bigr)}.
\end{equation}
If $\chi(0)=\pm 1$, then $\chi(t)=\pm 1$ and one finds $\Psi(t)=\pm
\Phi(t)$ for all times, where $\Psi(t)$ obeys
\begin{equation}
\frac{d\Psi}{dt} = p(1 \mp \sigma\Psi)\Psi, \label{apo1}
\end{equation}
whose solution is
\begin{equation}
{\Psi(t)}^{-1}= \pm\sigma + \exp(-pt)[{\Psi(0)}^{-1} \mp \sigma].
\end{equation}
Note that in these particular cases, only particles with velocity $+c$
(or $-c$) are present in the system at all times.

If $\chi(0) \not = 1$, one finds
\begin{eqnarray}
\lefteqn{\frac{d\Psi}{dt}
 =  p(1 - \sigma\Phi)\Psi
 =  p \Psi\,\times} \label{ap3} \\
&&  \biggl\{1
 -        \biggl[
                \frac{1+\chi(0)+[\chi(0)-1]\exp
                      \bigl(-c\int_0^t \Psi(\tau) d\tau\bigr)}
                     {1+\chi(0)-[\chi(0)-1]\exp
                      \bigl(-c\int_0^t \Psi(\tau) d\tau\bigr)}
          \biggr]\sigma \Psi
    \biggr\}. \nonumber
\end{eqnarray}
As $\vert \Psi(t)  \vert$ is a nondecreasing function of time, when $t \to \infty$, the square bracket in eq.~(\ref{ap3}) approaches $\pm 1$ depending on 
the  sign of $\chi(0)$. Thus, eq.~(\ref{ap3}) reduces to eq.~(\ref{apo1}).

}%end appendix
\end{multicols}
\centerline{\rule{25pc}{0.5pt}}
\begin{multicols}{2}

\end{multicols}

\begin{references}
\bibitem{EF}
        Y. Elskens and H. L. Frisch, Phys.\ Rev.\ A {\bf 31}, 3812 (1985).
\bibitem{KS}
        J. Krug and H. Spohn, Phys.\ Rev.\ A {\bf 38}, 4271 (1988).
\bibitem{BRL}
        E. Ben-Naim, S. Redner and F. Leyvraz, Phys.\ Rev.\ Lett.\ {\bf
        70}, 1890 (1993).
\bibitem{R}
        P. L. Krapivsky, S. Redner and F. Leyvraz, Phys.\ Rev.\ E {\bf
        51}, 3977 (1995).
\bibitem{jarek_uno}
        J. Piasecki, Phys.\ Rev.\ E {\bf 51}, 5535 (1995).
\bibitem{jarek_due}
        M. Droz, P.-A. Rey, L. Frachebourg and J. Piasecki,
        Phys.\ Rev.\ Lett.\ {\bf 75}, 160 (1995).
\bibitem{jarek_tre}
        M. Droz, P.-A. Rey, L. Frachebourg and J. Piasecki,
        Phys.\ Rev.\ E {\bf 51}, 5541 (1995).
\bibitem{jarek_four}
        J. Piasecki, P.-A. Rey and M. Droz, Physica A {\bf 229}, 515
        (1996).
\bibitem{jarek_five}
        P.-A. Rey, M. Droz and J. Piasecki, Eur.\ J. Phys.\ {\bf 18},
        213 (1997).
\bibitem{univer}
	P.-A. Rey, M. Droz and J. Piasecki, Phys.\ Rev.\ E {\bf
        57}, 138 (1998).
\bibitem{cardy_uno}
	J. L.~Cardy and U. C.~T\"auber, Phys.\ Rev.\ Lett.\ {\bf 77}, 4780
	(1996).
\bibitem{cardy_due}
	J. L.~Cardy and U. C.~T\"auber, J.\ Stat.\ Phys.\ {\bf 90}, 1
	(1998).
\bibitem{dp}
	P.~Grassberger and K.~Sundermeyer, Phys.\ Lett.\ {\bf 77 B}, 220
	(1978); J. L.~Cardy and R. L.~Sugar, J. Phys.\ A {\bf 13}, L423
	(1980);
\bibitem{bbgky}
	H. J. Kreutzer, {\it Nonequilibrium Thermodynamics and its
	Statistical Foundations}, Oxford Science Publication, (1981).
\bibitem{frach} L. Frachebourg and P. L. Krapivsky,  Phys.\ Rev.\ E {\bf
        55}, 252 (1997).
\end{references}
\end{document}